\documentclass[aps,prl,twocolumn,superscriptaddress,longbibliography]{revtex4-1}

\usepackage{graphicx}
\usepackage{bm}
\usepackage{hyperref}
\usepackage{todonotes}
\usepackage{color}
\usepackage{comment}
\usepackage{verbatim}
\usepackage{soul}

\def\LNO{LaNiO$_3$}
\def\LSMO{La$_{2/3}$Sr$_{1/3}$MnO$_3$}

\begin{document}

\title{Emergent $c$-axis magnetic helix in manganite-nickelate superlattices}

\author{G. Fabbris}
\email{gfabbris@anl.gov}
\affiliation{Department of Condensed Matter Physics and Materials Science, Brookhaven National Laboratory, Upton, New York 11973, USA}
\affiliation{Advanced Photon Source, Argonne National Laboratory, Argonne, Illinois 60439, USA}

\author{N. Jaouen}
\affiliation{Synchrotron SOLEIL, L'Orme des Merisiers, Saint-Aubin, BP 48, 91192 Gif-sur-Yvette, France}

\author{D. Meyers}
\affiliation{Department of Condensed Matter Physics and Materials Science, Brookhaven National Laboratory, Upton, New York 11973, USA}

\author{J. Feng}
\altaffiliation[Present address: ]{CAS Key Laboratory of Magnetic Materials and Devices, Ningbo Institute of Materials Technology and Engineering, Chinese Academy of Sciences, Zhejiang 315201 Ningbo, China}
\affiliation{Sorbonne Universit\'{e}, CNRS, Laboratoire de Chimie Physique-Mati\'{e}re et Rayonnement,UMR 7614, 4 place Jussieu, 75252 Paris Cedex 05, France}

\author{J. D. Hoffman}
\altaffiliation[Present address: ]{Department of Physics, Harvard University, Cambridge, MA 02138, USA}
\affiliation{Materials Science Division, Argonne National Laboratory, Argonne, Illinois 60439, USA}
\affiliation{Nanoscience and Technology Division, Argonne National Laboratory, Argonne, Illinois 60439, USA}

\author{R. Sutarto}
\affiliation{Canadian Light Source, Saskatoon, Saskatchewan S7N 2V3, Canada}

\author{S. G. Chiuzb\u{a}ian}
\affiliation{Sorbonne Universit\'{e}, CNRS, Laboratoire de Chimie Physique-Mati\'{e}re et Rayonnement,UMR 7614, 4 place Jussieu, 75252 Paris Cedex 05, France}
\affiliation{Synchrotron SOLEIL, L'Orme des Merisiers, Saint-Aubin, BP 48, 91192 Gif-sur-Yvette, France}

\author{A. Bhattacharya}
\affiliation{Materials Science Division, Argonne National Laboratory, Argonne, Illinois 60439, USA}
\affiliation{Nanoscience and Technology Division, Argonne National Laboratory, Argonne, Illinois 60439, USA}

\author{M. P. M. Dean}
\email{mdean@bnl.gov}
\affiliation{Department of Condensed Matter Physics and Materials Science, Brookhaven National Laboratory, Upton, New York 11973, USA}

\date{\today}

\begin{abstract}

The nature of the magnetic order in (\LSMO)$_9$/(\LNO)$_3$ superlattices is investigated using x-ray resonant magnetic reflectometry. We observe a new $c$-axis magnetic helix state in the (\LNO)$_3$ layers that had never been reported in nickelates, and which mediates the $\sim 130^{\circ}$ magnetic coupling between the ferromagnetic (\LSMO)$_9$ layers, illustrating the power of x-rays for discovering the magnetic state of complex oxide interfaces. Resonant inelastic x-ray scattering and x-ray absorption spectroscopy show that Ni-O ligand hole states from bulk \LNO\ are mostly filled due to interfacial electron transfer from Mn, driving the Ni orbitals closer to an atomic-like $3d^8$ configuration. We discuss the constraints imposed by this electronic configuration to the microscopic origin of the observed magnetic structure. The presence of a magnetic helix in (\LSMO)$_9$/(\LNO)$_3$ is crucial for modeling the potential spintronic functionality of this system and may be important for designing emergent magnetism in novel devices in general.

\end{abstract}

\maketitle

\begin{figure} [t]
\includegraphics{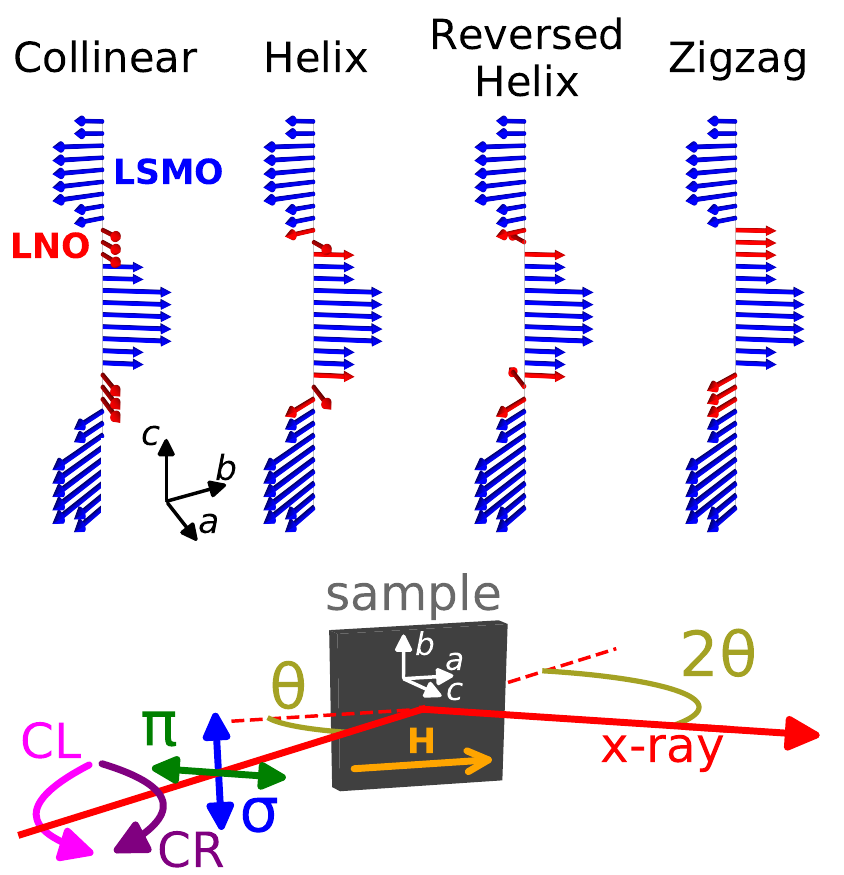}
\caption{Top: Possible magnetic structures of the \LNO{} layer in (LSMO)$_9$/(LNO)$_3$. Bottom: Experimental geometry used in both XRMR and RIXS measurements. XRMR was measured using circular and linear x-ray polarization with the magnetic field (H) applied along the sample's $a$-axis. RIXS was measured using linear x-ray polarization with $\theta$ and $2\theta$ fixed to $15^{\circ}$ and $85^{\circ}$, respectively.}
\label{schemes}
\end{figure}

Interfaces between complex oxide materials exhibit remarkably rich physics driven by the interplay of various types of charge, orbital and spin couplings \cite{Zubko2011,Bhattacharya2014,Hellman2017}. Particularly fascinating is their proven ability to host new types of emergent magnetic order that do not exist in either of the bulk constituents, representing a challenge to our understanding of electron correlations, as well as an opportunity for exploitation in spintronic devices \cite{Bhattacharya2014,Hellman2017,Hoffman2018}. In fact, complex magnetic phases are potentially common in transition metal oxide superlattices given the severe effects of interfacial symmetry breaking in localized $3d$ orbitals \cite{Banerjee2013,Li2014,Fust2016,Moetakef2012,Brinkman2007}, but scrutiny over the microscopic magnetic interactions at such interfaces is often constrained by limited direct evidence of the resulting magnetic structure. Superlattices composed of manganites and nickelates are a paradigmatic venue for efforts to discover and understand emergent interface magnetism, motivated by the complex magnetic order and phase diagram of its bulk constituents \cite{Zener1951,Torrance1992}. In fact, this system harbors fascinating phenomena, such as exchange bias and interfacial electronic reconstructions \cite{Nikolaev1999,Nikolaev2000,RojasSanchez2012,Gibert2012,Hoffman2013,Lee2013,Gibert2015,Piamonteze2015,Ning2015,Hoffman2016,Gibert2016,Kitamura2016,Zang2017,Flint2017a,Flint2017b}. Recently, a highly unusual magnetic coupling between ferromagnetic (LSMO)$_9$ layers was observed in $[001]$-oriented (\LSMO)$_9$/(\LNO)$_n$ [(LSMO)$_9$/(LNO)$_n$, $n = 1-9$], in which the coupling angle between (LSMO)$_9$ layers varies between zero and $130^{\circ}$ as a function of $n$ (Fig.~\ref{schemes}) \cite{Hoffman2016}. Despite the demonstrated ability of this system to be used as resistive memory devices \cite{Hoffman2018}, there is no experimental evidence for how or if the non-collinear magnetism of the (LSMO)$_9$ layers is mediated through (LNO)$_n$. Such lack of information critically hampers the ability to understand the physical mechanism driving the (LSMO)$_9$ magnetic coupling, which consequently inhibits the design of new superlattices with optimized magnetic properties.

In this work we focus on the (LSMO)$_9$/(LNO)$_3$ superlattice because it displays the largest magnetic coupling angle between the manganite layers ($\approx 130^{\circ}$ at 25~K \cite{Hoffman2016}). We probe its magnetic and electronic structure using resonant soft x-ray techniques. The magnetic structure of both (LSMO)$_9$ and (LNO)$_3$ are studied using Mn and Ni $L$-edge x-ray resonant magnetic reflectivity (XRMR). We find that the nickelate layers contain a magnetic helix in which the Ni moments are confined to the basal plane and rotate around the $c$-axis. Such magnetic structure is not present in any simple perovskite nickelate nor manganite. The XRMR ability to provide a precise description of the (LSMO)$_9$/(LNO)$_3$ magnetic ordering allows for a careful analysis of the magnetic interactions in this system. To this end, the nature of the Ni $3d$ and O $2p$ orbitals is probed through Ni $L_3$-edge resonant inelastic x-ray scattering (RIXS) and O $K$-edge x-ray absorption spectroscopy (XAS) measurements. The RIXS spectra are composed of well defined orbital excitations, which are reproduced by a model of localized Ni $3d^8$ orbitals under the effect of an octahedral crystal field. Even though the Ni $3d$ orbital is predominantly localized, O $K$-edge XAS show a small Ni $3d$ - O $2p$ ligand hole concentration, indicating that the NiO$_2$ planes are lightly hole-doped. While further work is needed to elucidate the magnetic couplings acting in this system, our analysis suggests that the (LSMO)$_9$/(LNO)$_3$ magnetic helix emerges from a combination between the strong pinning of Ni and Mn moments at the interfaces with a spin density wave (SDW) instability within the (LNO)$_3$ layers.

High-quality $14\times$[(LSMO)$_9$/(LNO)$_3$] superlattices were grown on $[001]$ single-crystalline SrTiO$_3$ substrates using molecular beam epitaxy as described previously \cite{Hoffman2016,supplemental}. XRMR measurements were performed at the Ni and Mn $L$-edges \cite{supplemental} at the REIXS beamline of the Canadian Light Source \cite{Hawthorn2011} and at the SEXTANTS beamline of the SOLEIL synchrotron \cite{Jaouen2004}. Data were collected using horizontal scattering geometry, as well as both linear ($\sigma$ and $\pi$) and circular [left (CL) and right (CR)] x-ray polarizations (Fig.~\ref{schemes}). Magnetic field was applied parallel to the sample surface and within the scattering plane using permanent magnets at REIXS and an electromagnet at SEXTANTS. In the former the data was collected at remanence after removing a 600~mT field, while in the later the sample was field cooled, and measured, under 1.2~mT. The temperature was kept at $\sim$25~K throughout the experiments. Data analysis was performed with the Dyna \cite{Elzo2012} and ReMagX \cite{Macke2014} softwares using the magnetic matrix formalism. Resonant optical constants were obtained by combining the tabulated values with averaged XAS and x-ray magnetic circular dichroism measurements \cite{Chantler1995,supplemental}

\begin{figure}[b]
\includegraphics{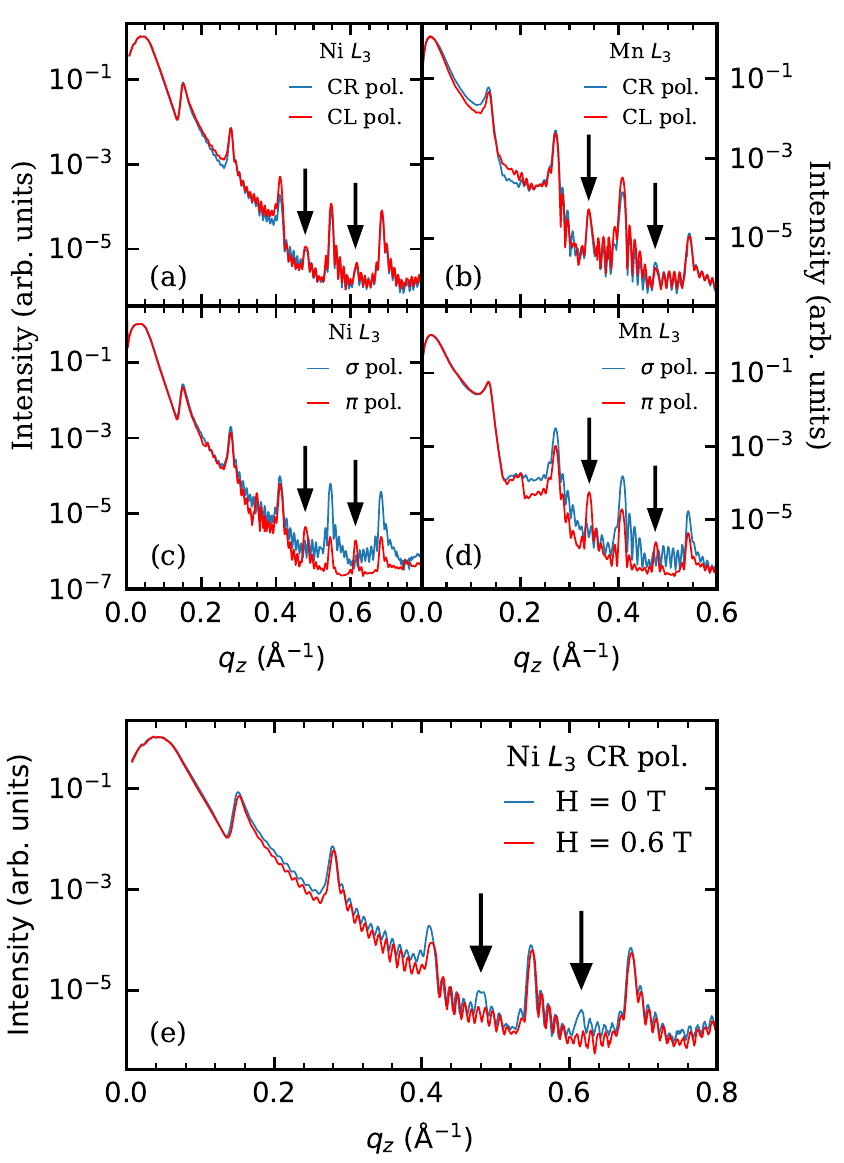}
\caption{(a)\&(c) Ni $L_3$-edge (853.6 eV) XRMR data using circular and linear polarization, respectively. (b)\&(d) Mn $L_3$-edge (641 eV) XRMR data using circular and linear polarization respectively. Data from both edges show half-order magnetic peaks (black arrows). This is consistent with a non-collinear magnetic structure between both LNO layers and LSMO layers as shown in Fig.~\ref{schemes}. (e) Field dependence of the Ni $L_3$ XRMR data. A field of 0.6~T is sufficient to force a ferromagnetic alignment between every layer, which suppresses the half-order peaks, consistent with previous results \cite{Hoffman2016}.}
\label{xrmr_93}
\end{figure}

Figure \ref{xrmr_93} displays a summary of the XRMR data collected on (LSMO)$_9$/(LNO)$_3$. The XRMR signal measured with circular light is primarily sensitive to components of the magnetization of each layer along the x-ray direction, while $\pi$ polarization probes the transverse magnetization, and $\sigma$ polarization is dominated by the lattice response \cite{Elzo2012,Macke2014}. It is thus clear in Figs. \ref{xrmr_93} (a) through (d) that magnetic peaks appear at $1/2$ order positions in (LSMO)$_9$/(LNO)$_3$ at both Mn and Ni edges indicating that the magnetic structure unit cell along the $c$-axis has twice the length of the chemical repetition. This observation is further demonstrated by the disappearance of such magnetic peaks when a field of 0.6~T is applied, which is sufficient to saturate the magnetic moments along the field [Fig.~\ref{xrmr_93}(e)] \cite{Hoffman2016}. While the Mn edge results are consistent with the previous report \cite{Hoffman2016}, the data at the Ni edge indicates that a non-collinear magnetic structure is also present in the (LNO)$_3$ layer. These results also demonstrate that the magnetic coupling within NiO$_2$ planes is not antiferromagnetic as this would lead to the absence of magnetic response in the CR/CL XRMR. Consequently, the magnetic structure of (LNO)$_3$ is different from the $(\frac{1}{4} \frac{1}{4} \frac{1}{4})$-type observed in [111]-grown LNO heterostructures and bulk $RE\mathrm{NiO_3}$ \cite{Torrance1992,Gibert2016,Note2018}. However, this in itself does not distinguish between other (LNO)$_3$ magnetic structures such as the zigzag and (reversed) helical models (Fig.~\ref{schemes}). 

\begin{figure}[t]
\includegraphics{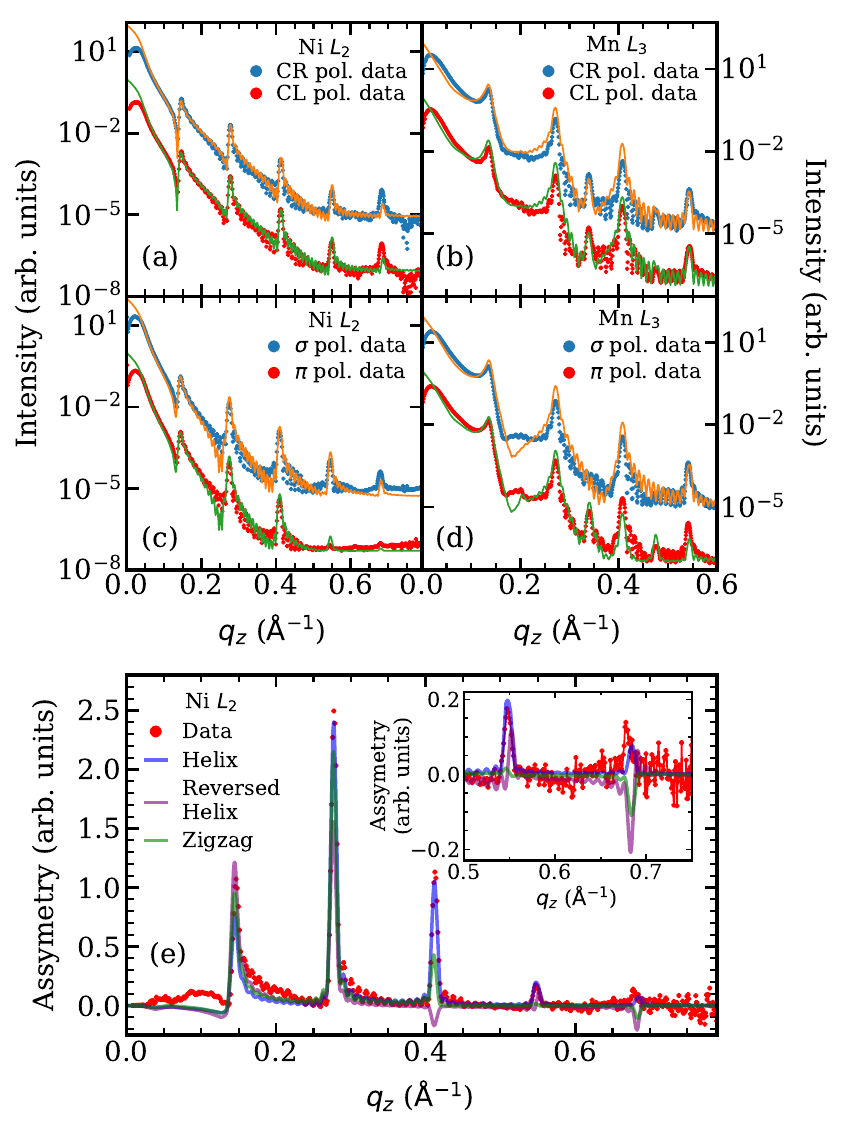}
\caption{Fits to the XRMR signal at the Ni $L_2$ (871.6 eV) [(a)\&(c)] and Mn $L_3$ [(b)\&(d)] edges in (LSMO)$_9$/(LNO)$_3$. Different polarization channels CR/CL or $\sigma$/$\pi$ are shown as red/blue points and are overlapped by orange/blue lines representing the fit. Different channels are offset from one another for clarity. Panel (e) displays the XRMR asymmetry \cite{assymetry}; the helix model is consistent with the presence of peaks at high $q_z$.}
\label{xrmr_fit_93}
\end{figure}

The (LNO)$_3$ magnetic structure is solved by modeling the XRMR data as shown in Fig.~\ref{xrmr_fit_93}. Note that fits were performed at the Ni $L_2$ edge because there is a substantial overlap between the La $M_4$ and Ni $L_3$  edges, which makes the XRMR data analysis very difficult. Despite its magnetic sensitivity, the XRMR signal is largely dominated by the lattice and charge responses [Fig.~\ref{xrmr_fit_93}(a)-(d)]. In fact, the disagreements between simulation and experimental data seen at the Mn $L_3$-edge are at least partially driven by the expected Mn$^{4+/3+}$ valence modulation within the (LSMO)$_9$ layer \cite{Hoffman2013,Zwiebler2015,Hoffman2016}. We can overcome this challenge by isolating the magnetic response using the XRMR circular polarization asymmetry shown in Fig.~\ref{xrmr_fit_93}(e) \cite{assymetry}. It is clear that only the helical model is able to reproduce all features in the data, particularly at high-$q_z$. Magnetic models with LNO spins away from the basal plane are not shown here but were also unable to reproduce the experimental data. Modeling of both Mn and Ni edges is consistent with an angle of $140\pm20^{\circ}$ between (LSMO)$_9$ magnetic moments, in agreement with neutron reflectivity \cite{Hoffman2016}. The (LNO)$_3$ $c$-axis magnetic helix result is surprising as such structures are uncommon in bulk transition metal oxides and have not previously been observed in nickelates. This demonstrates that in (LSMO)$_9$/(LNO)$_3$ the interfacial symmetry breaking and charge transfer drive an emergent form of magnetism, the underlying electronic structure of which we now explore.

\begin{figure}[b]
\includegraphics{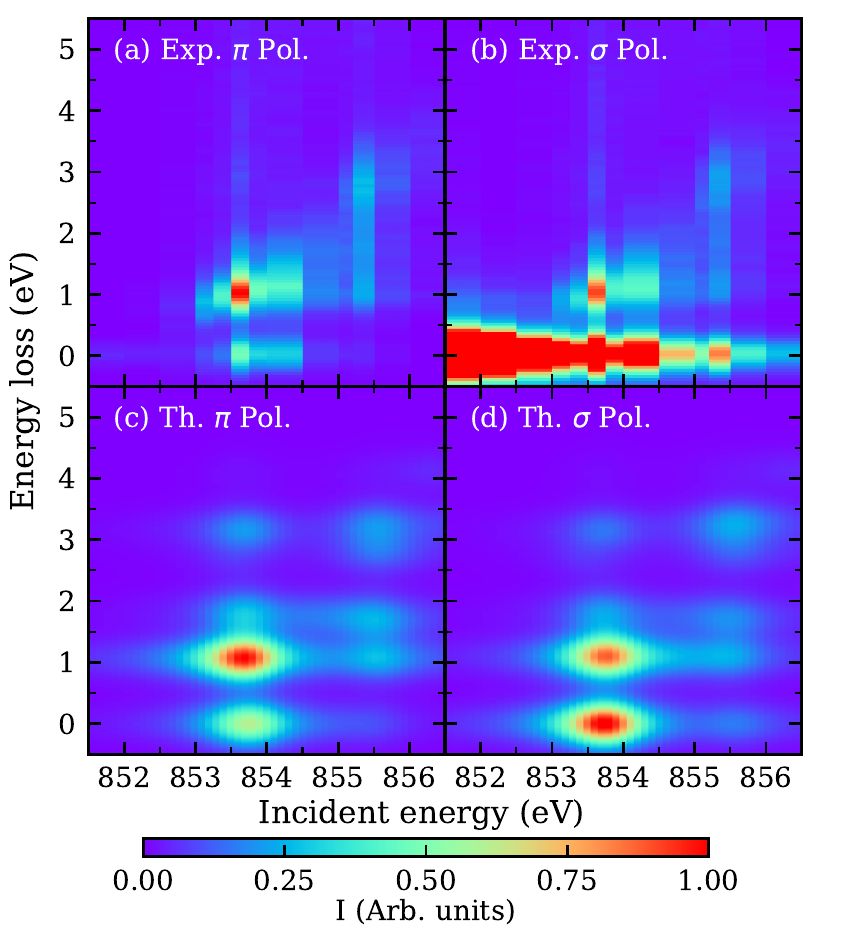}
\caption{Ni $L_3$-edge RIXS of (LSMO)$_9$/(LNO)$_3$. Data for $\pi$ and $\sigma$ polarized incident x-rays are displayed in panels (a)\&(b), respectively. Panels (c)\&(d) present simulations of the RIXS spectra for $\pi$ and $\sigma$ polarization using atomic calculations \cite{CowmanBook,deGrootBook,Fabbris2016}.}
\label{rixs_maps_93}
\end{figure}

We first address the character of the nickelate layer $3d$ orbital using Ni $L_3$-edge RIXS. Data was collected at 30~K using the AERHA end-station of the SEXTANTS beamline at the SOLEIL synchrotron \cite{Chiuzbaian2014}. The experimental scattering geometry is detailed in Fig.~\ref{schemes}. Figure \ref{rixs_maps_93} shows the RIXS spectra as a function of incident x-ray energy and linear polarization. The experimental data [Fig.~\ref{rixs_maps_93} (a)\&(b)] is composed of narrow peaks that are consistent with localized orbital excitations of $3d^8$ (Ni$^{2+}$) ions \cite{Ghiringhelli2005,Ghiringhelli2009,Fabbris2016,Fabbris2017}; a distinctively different broad diagonal feature is seen in RIXS spectra of bulk $RE$NiO$_3$ and hole-doped $\mathrm{La_{2-x}Sr_xNiO_4}$ (for $x \geq 1/3$) \cite{Bisogni2016,Fabbris2016,Fabbris2017}, in which the significant amount of extra holes in the NiO$_2$ plane populate the hybridized Ni $3d$ - O $2p$ ligand hole states. This result is consistent with Ni $L_2$-edge XAS data \cite{Hoffman2016}. The RIXS spectra was simulated using the Kramers-Heisenberg equation combined with atomic multiplet calculations performed with the Cowan and RACER codes \cite{CowmanBook,deGrootBook,Fabbris2016}. These simulations [Fig.~\ref{rixs_maps_93} (c)\&(d)] nicely reproduce the observed signal \cite{supplemental}, confirming that these are orbital $dd$ excitations, thus demonstrating the localized character of the Ni $3d$ state in (LSMO)$_9$/(LNO)$_3$. Furthermore, the excellent agreement between experimental data and simulation implies high-spin ($S=1$) Ni $3d^8$ orbitals under a nearly octahedral crystal field with 10$D_q = 1.1$~eV and $\Delta e_g = 0.05$~eV \footnote{Best agreement with the experimental data was obtained with $\Delta t_{2g} = 0$ and reduced Slater-Condon parameters: $F_{dd} = 70\%$, $F_{pd} = 85\%$ and $G_{pd} = 70\%$. We point out that the $\Delta e_{g} = 0.05$~eV found here implies that $\Delta t_{2g}$ is likely finite. However, it also suggests a $\Delta t_{2g} < 0.01$~eV, which is smaller our sensitivity.}.

\begin{figure}[t]
\includegraphics{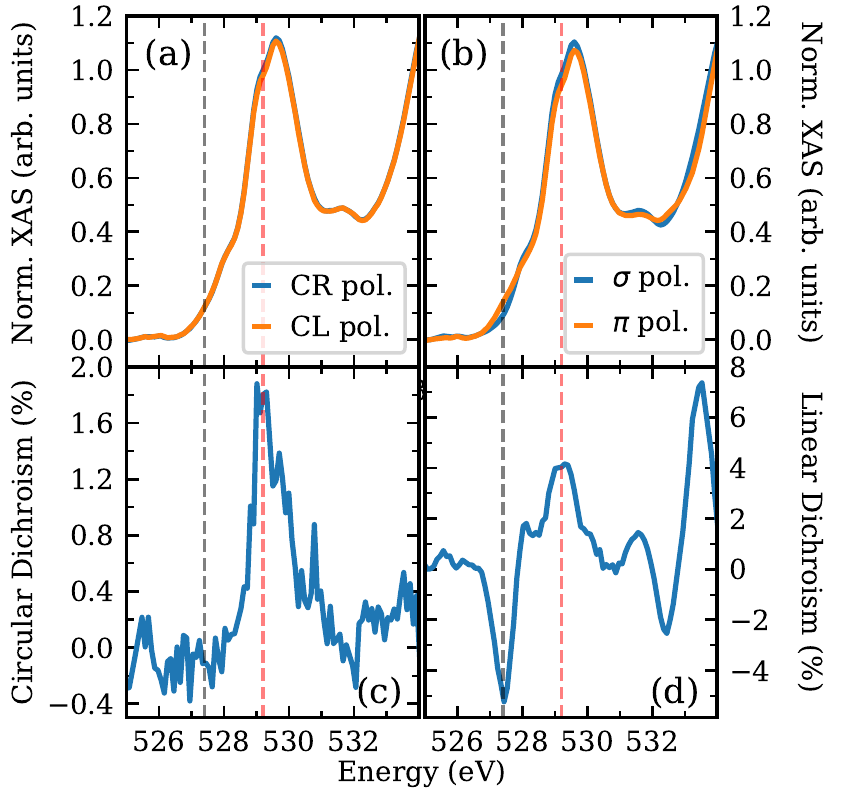}
\caption{(LSMO)$_9$/(LNO)$_3$ O $K$-edge XAS data collected at 15$^{\circ}$ incident angle. (a)\&(b) shows the pre-edge features of the XAS collected with circular and linear x-rays, respectively. The circular and linear dichroism are plotted in panels (c)\&(d), respectively. While signal from hybridized Mn-O orbitals dominate the pre-edge features (vertical red dashed line), a signal that is consistent with Ni-O hybridization is seen around 527.4 eV (grey dashed line).}
\label{ok_xas}
\end{figure}

While the RIXS results establish a predominantly localized Ni $3d$ orbital, it has minimal sensitivity to the potential presence of a small amount of Ni $3d$ - O $2p$ ligand holes. Figure \ref{ok_xas} displays the pre-edge of the O $K$-edge XAS data collected in total electron yield mode at the REIXS beamline of the Canadian Light Source. The x-rays were incident at $15^{\circ}$ and the sample was field cooled under 0.6~T, but the magnetic field was removed prior to the data collection. The pre-edge is dominated by features related to Mn-O and Ni-O ligand holes. Comparing this data with the literature \cite{Alonso2007,Cui2015}, it is clear that the Mn-O states dominate the signal. This result is consistent with the large response expected for Mn $3d^4$/$3d^3$ (Mn$^{3+/4+}$) and small or absent pre-edge structure for Ni $3d^8$ (Ni$^{2+}$) \cite{Kuiper1995,Pellegrin1996,Cao2016}. It is also consistent with the position of the magnetic circular dichroism signal, which is expected to be dominated by LSMO [Fig.~\ref{ok_xas}(a)\&(c)]. However, a small pre-edge shoulder is seen at an energy that is consistent with it originating in the LNO layer (grey dashed line). Interestingly this feature displays a marked linear dichroism, implying an anisotropic ligand hole state in which the more holes populate the $3d_{3z^2-r^2}2p_{z}$ states in detriment of $3d_{x^2-y^2}2p_{\sigma}$. This result is consistent with the tensile strain applied by the SrTiO$_3$ substrate and is in-line with the concept that strain primarily controls the Ni-O hybridization in LNO heterostructures \cite{Parragh2013,Grisolia2016,Fabbris2016}.

We now discuss the consequences of the observed electronic configuration to the potential mechanisms driving the complex magnetic structure of (LSMO)$_9$/(LNO)$_3$. We start by noting that non-collinear magnetic order in transition metal oxides may derive from the Dzyaloshinskii-Moriya interaction \cite{Dzyaloshinsky1958,Moriya1960}, but this interaction is inconsistent with the symmetry of the (LNO)$_3$ magnetic structure \cite{Hoffman2016}. The interfacial electronic reconstruction in (LSMO)$_9$/(LNO)$_3$ induces magnetic frustration in the nickelate layer since the antiferromagnetic Ni$^{2+}$-O-Ni$^{2+}$ exchange energy is expected to be similar to the dominant ferromagnetic Ni$^{2+}$-O-Mn$^{4+}$ \cite{Nakajima1993,Haskel2011,Lee2013,Fabbris2017}. However, while we cannot exclude frustration as the sole reason for the observed spiral, if we only consider first neighbor interactions, such frustration would arguably drive an antiferromagnetic coupling structure that resembles the reversed helical model (see Fig.~\ref{schemes}) since the adjacent NiO$_2$ planes would favor an antiferromagnetic coupling. Additionally, it is difficult to reconcile the magnetic structure LNO thickness dependence with a magnetic frustration model \cite{Hoffman2016}. Emergent magnetic phenomena are often seen in systems that, much like (LSMO)$_9$/(LNO)$_3$, have $3d$-$2p$ ligand holes (negative charge transfer systems \cite{Khomskii2001}), including certain cuprates \cite{Mizokawa1991}, hole-doped $\mathrm{La_2NiO_4}$ \cite{Pellegrin1996,Fabbris2017}, $RE$NiO$_3$ \cite{Bisogni2016}, and CrO$_2$ \cite{Korotin1998}. In fact, it has been argued that double-exchange and superexchange can act together to create non-collinear canted structures in such systems \cite{DeGennes1960,Mostovoy2005,Santos2011}, but it is not entirely clear how such canting would actually lead to a helix. Finally, we note that the electronic structure of (LNO)$_3$ closely mimics that of lightly hole-doped $\mathrm{La_2NiO_4}$ \cite{Pellegrin1996,Kuiper1998,Fabbris2017}, which displays stripe-like spin density wave instabilities \cite{Tranquada1994,Sachan1995}, which in turn are known to foster non-collinear magnetic structures \cite{Shraiman1988,Zachar1998}. Therefore, it appears that the (LNO)$_3$ magnetic helix emerges from the coupling of such SDW instability with the interfacial Ni-Mn interaction. 

In conclusion, the presence of an emergent magnetic helical structure in the [\LSMO]$_9$/[\LNO]$_3$ superlattice is demonstrated through XRMR measurements. Such $c$-axis magnetic helix has not been shown to occur in any other nickelate system. Ni $L_3$-edge RIXS and O $K$-edge XAS experiments show that interfacial charge transfer results in NiO$_2$ planes with lightly hole-doped Ni $3d^8$ atoms. We argue that such electronic structure likely drives SDW instabilities that may drive the helical magnetic structure, but further work is required to verify our arguments. Besides theoretical investigations, we believe that the role of octahedral rotations in this system need to be addressed as these are known to relevant to magnetic exchange. Nevertheless, the prospect of engineering spin density waves through interfaces to create novel magnetism is tantalizing, and, in principle, such approach should applicable to the various families of negative charge transfer transition metal oxides \cite{Khomskii2001}.

\begin{acknowledgments}
We thank Dr. Daniel Haskel for insightful discussions and Dr. Alessandro Nicolaou for support during the RIXS measurements at the SEXTANTS beamline. This material is based upon work supported by the U.S.\ Department of Energy, Office of Basic Energy Sciences, Early Career Award Program under Award Number 1047478. Work at Brookhaven National Laboratory is supported by the U.S. Department of Energy, Office of Science, Office of Basic Energy Sciences, under Contract No. DE-SC00112704. Work at Argonne is supported by the U.S. Department of Energy, Office of Science, under contract No. DE-AC-02-06CH11357. Part of the research described in this paper was performed at the Canadian Light Source, which is supported by the Canada Foundation for Innovation, Natural Sciences and Engineering Research Council of Canada, the University of Saskatchewan, the Government of Saskatchewan, Western Economic Diversification Canada, the National Research Council Canada, and the Canadian Institutes of Health Research. We acknowledge SOLEIL for provision of synchrotron radiation facilities. S.G.C. acknowledges the support of the Agence Nationale de la Recherche (ANR), under Grant No. ANR-05-NANO-074 (HR-RXRS).
\end{acknowledgments}

\bibliography{refs}

\end{document}